\documentclass[12pt]{article}

\usepackage{amssymb}

\usepackage{verbatim}

\def\bS {{\mathbb{S}}}
\def\bR {{\mathbb{R}}}
\def\bN {{\mathbb{N}}}
\def\bC {{\mathbb{C}}}

\def\bZ {{\mathbb{Z}}}

\def\cC {{\mathcal C}} 

\def\Di {\displaystyle}

\def\MR {\mathrm }

\def\sp {\mathrm {sp}}
\def\mb {{\bf b}} 
\def\mbt {{\widetilde {\bf b}}}
\def\mM {{\bf M}}

\makeatletter
\@addtoreset{equation}{section}

\makeatother

\newtheorem{theorem}{Theorem}[section]
\newtheorem{lemma}[theorem]{Lemma}

\newtheorem{remark}[theorem]{Remark}

\begin{document}

\bibliographystyle{plain}

\begin{center}
{\Large \bf {Eigenvalues of Laplacian with constant magnetic field on noncompact hyperbolic surfaces with finite area}}
\end{center}

\vskip 0.5cm

\centerline{\bf { Abderemane MORAME$^{1}$
 and  Fran{\c c}oise TRUC$^{2}$}}

{\it {$^{1}$ Universit\'e de Nantes,
Facult\'e des Sciences,  Dpt. Math\'ematiques, \\
UMR 6629 du CNRS, B.P. 99208, 44322 Nantes Cedex 3, (FRANCE), \\
E.Mail: morame@math.univ-nantes.fr}}

{\it {$^{2}$ Universit\'e de Grenoble I, Institut Fourier,\\
            UMR 5582 CNRS-UJF,
            B.P. 74,\\
 38402 St Martin d'H\`eres Cedex, (France), \\
E.Mail: Francoise.Truc@ujf-grenoble.fr }}

\begin{abstract}
We consider a  magnetic Laplacian  
 $-\Delta_A=(id+A)^\star (id+A)$\\  
 on a noncompact hyperbolic surface  $\mM  $ with finite area. 
 $A$ is a real one-form and 
  the magnetic field $dA$ is constant in each cusp. 
When the harmonic component of $A$ satifies some quantified 
condition, 
 the spectrum of $-\Delta_A$ is discrete. In this case  
we prove that   the counting function 
  of the eigenvalues of $-\Delta_{A}$ satisfies the classical Weyl formula, even when  $dA=0. $ 
\end{abstract}

\section{Introduction}

We consider  a smooth, connected, complete and oriented Riemannian surface 
$(\mM, g)$ and  a smooth, real 
 one-form $A$ on $\mM  .$  We  define the magnetic Laplacian, 
 the Bochner  Laplacian 
\begin{equation}\label{DeA} 
- \Delta_A\; =\; (i\ d + A)^\star (i\ d +A)\; ,
 \end{equation}
 $$
 ( \; 
 (i\ d+A)u=i\ du+uA\; , \ \forall \; u\; \in \; C^\infty_0(\mM ; \bC )\; .$$
 
The magnetic field is the exact two-form $\ \rho_B\; =\; dA\; .$

If $dm\; $ is the Riemannian measure on $\mM \; ,$ then 
\begin{equation}\label{DeMb} 
\rho_B\; =\; \mbt \; dm\; , \quad \MR{with} \quad \mbt \; \in \; 
C^\infty (\mM ;\bR )\; . 
\end{equation} 
The magnetic intensity is $\ \mb\; =\; |\mbt |\; .$ 

It is well known, (see \cite{Shu}  ), that $-\Delta_A$ has a unique 
self-adjoint extension on $L^2(\mM )\; ,$ containing in its domain 
$C_0^\infty (\mM ;\bC )\; ,$ the space of 
smooth and compactly supported functions. 
The spectrum of $-\Delta_A$ is gauge invariant~: 
for any $f\in C^1(\mM ; \bR )\; ,\ -\Delta_A$ 
and $-\Delta_{A+df}$ are unitarily equivalent, hence they have the same spectrum.

We are interested in constant magnetic fields on  $\mM \; $ in the case when
 $(\mM ,\; g)\; $ is a  non-compact geometrically finite hyperbolic surface of finite area; 
  (see \cite{Per} or \cite{Bor} for the definition 
and the related references). More precisely 
\begin{equation}\label{hyM} 
\mM \; =\; \bigcup_{j=0}^{J}M_j  \;  
\end{equation} 
where the $M_j$  are open sets of \mM , such 
that the closure of $M_0$ is compact, and  (when $J\geq 1)\; $ the other 
$M_j$ are cuspidal ends of \mM .  

This means that, for any $j,\ 1\leq j\leq J\; ,$ there exist strictly positive constants $a_j\ \MR{and}\ L_j$ such that $M_j\; $ is isometric to 
$\bS \times ] a_j^2, +\infty [\; ,$ equipped with the metric 
\begin{equation}\label{gCups} 
 ds_j^2\; =\; y^{-2} (\ L_j^2 \ d\theta^2\; +\; dy^2\ )\; ;
\end{equation} 
$(\Di  \bS \; =\; \bS^1\; $   is the unit circle
 and 
 $ M_j\cap M_k\;  =\; \emptyset $ if $j\neq k \; )\; .$ \\   
Let us choose some $z_0\; \in \; M_0\; $ and let us define 
\begin{equation}\label{Dd} 
d\; :\; \mM \; \to \; \bR_+\; ;\quad d(z)\; =\; d_g(z,z_0)\; ; 
\end{equation} 
$d_g (\; .\; ,\; .\; )\; $ denotes the distance with respect to the metric
 $g$.
 
For any $\Di b \; \in \; \bR^{J}\; ,$ 
there exists a one-form $\ A\; ,$ such that the corresponding magnetic field 
$dA\; $ satisfies
\begin{equation}\label{asymptC} 
dA\; =\; \mbt (z) dm\; \quad \MR{with}\quad  
\mbt (z)\; =\; b_j\ \forall\; z\; \in \;  M_j \; .
\end{equation}

The following statement on the essential spectrum is proven in \cite{Mo-Tr1}~:
\begin{theorem}\label{ThC} 
 Assume (\ref{hyM}) and (\ref{asymptC}). 
 Then for any $j\; ,\ 1\leq j\leq J\; $ 
and for any $\; z\; \in \; M_j\; $ 
there exists a unique closed curve through $\Di \; z\; ,\   
 \cC_{j,z}\; $ in $\ (M_j, \ g)\; ,$ 
not contractible and  with zero $g-$curvature. ($ \cC_{j,z}$
is called an horocycle of $ M_j $ ).
The following limit 
exists and is finite: 
\begin{equation}\label{Aj} 
[A]_{M_j}\; =\; \lim_{d(z)\to +\infty} \; \int_{\cC_{j,z}} 
A\; . 
\end{equation} 

If $J^A\; =\; \{ j\in \bN \; , \ 1\leq j\leq J\ s.t.\ 
[A]_{M_j}\in 2\pi \bZ\; \} \; \neq \; \emptyset \; ,$ then 
\begin{equation}\label{CThC} 
\sp_{ess} (-\Delta_A)\; =\; 
[\frac{1}{4}+\min_{j\in J^A} b_j^2 \;  , \; +\infty [\; . 
\end{equation} 

If  $J^A\; =\; \emptyset \; ,$ then 
$\ \sp_{ess} (-\Delta_A)\; =\; \emptyset \; :$\\ 
$-\Delta_A\; $ has purely discrete spectrum, (its resolvent 
is compact). 

\end{theorem}

When the magnetic Laplacian  $-\Delta_A\; $ has purely discrete spectrum, it is called  a magnetic bottle, (see \cite{Col2}).

If $A=df +A^H + A^\delta $ is the Hodge decomposition 
of $A$ with $A^H$ harmonic, $(dA^H=0$ and $d^\star A^H=0\; )\; ,$ 
then $\forall \; j\; ,\ [A]_{M_j}=[A^H]_{M_j}\; ,$ 
so  the discreteness of the spectrum of $-\Delta_A$ depends only 
on the harmonic component of $A\; .$ 
So one can see the case $J^A=\emptyset \; $ as 
an Aharonov-Bohm phenomenon \cite{Ah-Bo}, 
a situation where the magnetic field $dA$ is not sufficicient 
to describe $ -\Delta_A\; $ and the use of the magnetic potential 
 $A$ is essential~: 
we can have magnetic bottle with null intensity.

\section{The Weyl formula in the case  of finite area with a non-integer 
class one-form }

Here we are interested in the pure point part of the spectrum. 
 We assume that 
 $J^A\; =\; \emptyset \; ,$ 
then 
 the spectrum of $-\Delta_A\; $ 
is discrete. 
In this case,   we denote by $(\lambda_j)_j$ the increasing sequence of eigenvalues 
of $-\Delta_A\; ,$ (each eigenvalue is repeated according to its multiplicity). 
 Let 
\begin{equation}\label{DeNL} 
N(\lambda,-\Delta_A)\; =\; \sum_{\lambda_j < \lambda } 1\; .
\end{equation} 

We will show that the asymptotic behavior of $\; N(\lambda )$ 
is  given by the Weyl formula~: 
\begin{theorem}\label{ThD} 
Consider   a geometrically finite hyperbolic surface $(\mM ,\; g)\; $
of finite area,   
and assume (\ref{asymptC}) with    $J^A\; =\; \emptyset \; , $ 
 (see (\ref{Aj} for the definition). 

Then \begin{equation}\label{De} 
N(\lambda, -\Delta_A)\; =\; \lambda \frac{|\mM|}{4\pi}
\; + \; \bf{O}(\sqrt{\lambda} \ln \lambda )\; .
\end{equation} 

\end{theorem}

\begin{remark}As $J^A$ depends only on the harmonic component 
of $A\; ,\ J^A$ is not empty when $\mM $ is simply connected. 
In \cite{Go-Mo} there are some results close to Theorem \ref{ThD}, 
but for simply connected manifolds. 

The cases where the magnetic field prevails were studied 
in \cite{Mo-Tr1} and in \cite{Mo-Tr2}. 
\end{remark}
\noindent 
{\bf Proof of Theorem \ref{ThD}. } 
 Any constant depending only on the $b_j$ and on 
  $\displaystyle \min_{1\leq j\leq J}\inf_{k\in \bZ } |[A]_{M_j} -2k\pi |$  will be denoted invariably 
 $C\; .$

 Consider a cusp $M =M_j=\; \bS \times ]\alpha^2 , +\infty [\; $ equipped with the metric
 
$ ds^2\; =\; L^2e^{-2t} d\theta^2 \; +\; dt^2\; $ for some $\alpha >0$ and $ L > 0 \; .$ \\
Let us denote by $-\Delta_{A}^{M}$ the Dirichlet operator on $M\; ,$ 
associated to  $-\Delta_{A}\; $ . The first step will be to prove that
\begin{equation}\label{Dem} 
N(\lambda, -\Delta_A^{M})\; =\; \lambda \frac{|M|}{4\pi}
\; + \; \bf{O}(\sqrt{\lambda} \ln \lambda )\; .
\end{equation}
Since $-\Delta_A^{M}$ and $-\Delta_{A+d\varphi +k d\theta}^{M}$ are gauge equivalent for any $\varphi \in C^\infty (\overline{\mM} ;\bR )\;$
and any $k \in \bZ$,
 we can assume that 
$$-\Delta_{A}^{M}= 
L^{-2} e^{2t}(D_\theta -A_1)^2 
+  D_t^2 +\frac{1}{4}\; ,\quad \MR{with} 
\quad A_1=-\xi \pm bLe^{-t}\; ,\ \xi \in ]0,1[\; ,$$ 
$(b=b_j\; ,\ 2\pi \xi -[A]_M\; \in \; 2\pi \bZ )\; .$ Then we get that  
$$\sp (-\Delta_{A}^{M})=\bigcup_{\ell \in \bZ} 
\sp (P_\ell )\; ;\ P_\ell 
= D_t^2 + \frac{1}{4} + \left ( e^t \frac{(\ell + \xi )}{L} \pm b 
\right )^2 \; ,$$ 
for the Dirichlet condition on $L^2(I; dt)\; ;\ I=]\alpha^2, +\infty [\; .$ 
This implies that 
 \begin{equation}\label{Del} 
N(\lambda, -\Delta_{A}^{M})\; =\ \sum_{\ell \in \bZ} N(\lambda ,P_\ell) = \sum_{\ell \in X_\lambda} N(\lambda ,P_\ell)
\end{equation}
with $\Di X_\lambda = \{ \ell\ /\ e^{\alpha^2} \frac{|\ell + \xi |}{L} 
< \sqrt{\lambda - 1/4} - b \ \}\ .$ \\ 
 Denoting by $Q_\ell$ the Dirichlet operator on $I$ associated to 
$$\ Q_\ell 
= D_t^2 + \frac{1}{4} + \frac{(\ell +\xi )^2}{L^2} e^{2t} \; ,$$ 
we easily get that
\begin{equation}\label{compar2} 
Q_\ell-C\sqrt{Q_\ell}  \; \leq \; P_\ell \; \leq \; Q_\ell+C\sqrt{Q_\ell} \ .
\end{equation} 

Therefore one can find a constant $C(b)\; ,$ depending only on $b\; ,$ such that, for any $\lambda >> 1+C(b)\; ,$  
\begin{equation}\label{compar1} 
N(\lambda -\sqrt{\lambda}C(b) , Q_\ell) \; \leq \; N(\lambda ,P_\ell) \; \leq \; 
N(\lambda +\sqrt{\lambda}C(b) , Q_\ell) \; .
\end{equation} 
Following Titchmarsh's method ( \cite{Tit}, Theorem 7.4) we establish  the following bounds 
\begin{lemma}\label{lemN2} 
There exists  $C > 1 $ so that for any 
$\mu  >> 1\; $ and any $\ell \in X_\mu   \; ,$ 
\begin{equation}\label{tit}  w_\ell (\mu ) -\pi
\; \leq \; \pi N(\mu -\frac{1}{4} ,Q_\ell) \; \leq \; 
w_\ell (\mu ) + \frac{1}{12} \ln \mu + C
 \; ,
 \end{equation} 
with\begin{equation}\label{defwll} 
w_\ell (\mu )\; =\; 
\int_{\alpha^2}^{+\infty}  \left [ 
\mu   - \frac{(\ell +\xi )^2}{L^2} e^{2t}\right ]_{+}^{1/2} dt 
\end{equation}
$$ =\; 
\int_{\alpha^2}^{T_{\mu ,L}} \left [ 
\mu   - \frac{(\ell +\xi )^2}{L^2} e^{2t}\right ]_{+}^{1/2} dt \; ;$$
$\Di  (e^{T_{\mu ,L}}=L\sqrt{\mu}/(\inf_{k\in \bZ} |\xi -k|)\; )\; . 
$
\end{lemma} 

\noindent 
{\bf Proof of Lemma \ref{lemN2} } 

The lower bound is easily obtained (see \cite{Tit}, Formula 7.1.2 p 143) so
we focus on the upper bound.

Let us define $V_\ell =\frac{(\ell +\xi )^2}{L^2} e^{2t}$ and denote by $\phi_{\mu}^\ell$ a solution of $Q_\ell \phi= (\mu-\frac{1}{4}) \phi$.
 Consider $x_\ell$  and $y_\ell$ so that $V_\ell (x_\ell) = \mu$ and $V_\ell (y_\ell) = \nu$, for a given 
$0<\nu<\mu$ to be determined later.
We denote by $m$ the number of zeros of $\phi_{\mu}^\ell$ on $]\alpha^2,y_\ell[$.
Recall that the number $n$ of zeros of $\phi_{\mu}^\ell$ on $]\alpha^2,x_\ell[$ is equal to $ N(\mu-\frac{1}{4}  ,Q_\ell) $.

\noindent
 Applying Lemma 7.3 p 146 in \cite{Tit} we deduce that
$$m\pi = \int_{\alpha^2}^{y_\ell}  \left [ 
\mu   -V_\ell\right ]^{1/2} dt + R_\ell$$
with $R_\ell =\frac{1}{4} \ln (\mu -V_\ell(\alpha^2))-\frac{1}{4}\ln (\mu -V_\ell(y_\ell))+\pi$,
 hence
$$|n\pi -\int_{\alpha^2}^{x_\ell}  \left [ 
\mu   -V_\ell\right ]^{1/2} dt |\leq (x_\ell -y_\ell) (\mu -\nu)^{1/2} + R_\ell +(n-m)\pi$$
According to the  Sturm comparison theorem (\cite{Tit},  p 107-108), we have
$$(n-m)\pi\leq (x_\ell -y_\ell) (\mu -\nu)^{1/2}$$

and
$$|n\pi -\int_{\alpha^2}^{x_\ell}  \left [ 
\mu   -V_\ell\right ]^{1/2} dt |\leq \ln(\frac{\mu}{\nu})(\mu -\nu)^{1/2} +\frac{1}{4} \ln \mu -\frac{1}{4}\ln (\mu -\nu)+2\pi$$
Now taking $\nu = \mu-\mu^{2/3}$ we get the desired estimate.

 In view of (\ref{Del}) we now compute $\sum_{\ell \in \bZ}
w_\ell (\mu ) \ .$
We first get the following
\begin{lemma}\label{lemN3} 
There exists  $C> 1  \; $  such that, for any 
$\ \mu   >> 1\; $ and any $t\; \in \; [\alpha ^2 , T_{\mu,L}]  \; ,$ 
$$\left | \;  \int_\bR \left [ \mu -\frac{(x+\xi )^2}{L^2}e^{2t} 
\right ]_{+}^{1/2} dx \; - \; \sum_{\ell \in \bZ} 
\left [ 
\mu   - \frac{(\ell +\xi )^2}{L^2} e^{2t}\right ]_{+}^{1/2} \; \right |  
\; \leq \; C (\sqrt{\mu}  + \frac{e^t}{L})\; .
$$ 
\end{lemma} 
This leads to
\begin{lemma}\label{lemN4} 
There exists  $C > 1  \; $  such that, for any 
$\ \mu   >> 1\; , $  
$$\left | \;  \int_{\alpha^2}^{
T_{\mu ,L}}\int_\bR \left [ \mu -\frac{(x+\xi )^2}{L^2}e^{2t} 
\right ]_{+}^{1/2} dx dt  \; - \; \sum_{\ell \in \bZ} 
w_\ell (\mu ) 
 \; \right |  
\; \leq \; C \sqrt{\mu}  \ln \mu \; .
$$ 
\end{lemma} 
We now compute the integral in the left-hand side.

\noindent
Making the change of variables 
$y^2=\frac{(x+\xi )^2}{L^2\mu}e^{2t} $ we obtain that it is equal to\\
$\mu L \int_{\alpha^2}^{
T_{\mu ,L}} e^{-t} dt
 \int_\bR \left [ 1 -x^2 
\right ]_{+}^{1/2} dx ,
\ $ so we
get
\begin{lemma}\label{lemN5} 
There exists  $C > 1  \; $  such that, for any 
$\ \mu   >> 1\; ,$  
$$\left | \;  \int_{\alpha^2}^{T_{\mu,L}} \int_\bR \left [ \mu -\frac{(x+\xi )^2}{L^2}e^{2t} 
\right ]_{+}^{1/2} dx dt  \; -  \mu Le^{-\alpha^2} 
 \int_\bR \left [ 1 -x^2 
\right ]_{+}^{1/2} dx  
 \; \right |  
\; \leq \; C \sqrt{\mu } \; .
$$ 
\end{lemma}

\noindent
Noticing that $|M|= 2\pi L e^{-\alpha^2}$ and using Lemmas  \ref{lemN4} and \ref{lemN5} we have
 
\begin{lemma}\label{Ncla}

$$\frac{1}{\pi} \sum_{\ell} \in  
w_\ell (\mu ) \; =\;
\frac{|M|}{4\pi }\mu \; +\; {\bf O}(\sqrt\mu \ln \mu )\; ,
\quad {\rm as}  \quad \mu\to +\infty \; .$$

\end{lemma}

In view of (\ref{Del}),(\ref{compar1}) and (\ref{tit}) 
Lemma \ref{Ncla} ends  the proof of formula (\ref{Dem}). 

Now it remains to consider the whole surface $\mM$. 

We have~:
$\Di \ 
\mM \; =\; \left (\bigcup_{j=0}^{J}M_j\right )  \;  $\\ 
where the $M_j$  are open sets of $\mM ,$ such 
that the closure of $M_0$ is compact, and  the other 
$M_j$ are cuspidal ends of $\mM $ and \\ 
$M_j \cap M_k =\emptyset \; ,$ if $j\neq k\; .$ We denote 
$\Di \; M_0^0=\mM \setminus (\bigcup_{j=1}^{J}\overline{M_j} )\; ,$ 
then 
\begin{equation}\label{partit} 
\mM \; =\; \overline{M_0^0}\bigcup  \left (\bigcup_{j=1}^{J}\overline{M_j}\right ) \ .
\end{equation} 
Let us denote respectively by $-\Delta_{A,D}^{\Omega}$ and by $-\Delta_{A,N}^{\Omega}$ the Dirichlet 
operator and the Neumann-like operator on an open set 
$\Omega \; $ of $\mM $  
associated to  $-\Delta_{A}\; .$ 
 \\  
The minimax principle and (\ref{partit}) imply that
 \begin{equation}\label{Delll} 
N(\lambda, -\Delta_{A,D}^{M_0^0}) + \sum _{1\leq j\leq J} N(\lambda, -\Delta_{A,D}^{M_j}) \leq N(\lambda, -\Delta_{A})  
\end{equation}
$$\leq N(\lambda, -\Delta_{A,N}^{M_0^0}) + \sum _{1\leq j\leq J} N(\lambda, -\Delta_{A,N}^{M_j})$$
\indent 
The Weyl formula with remainder, (see  \cite{Hor} for Dirichlet boundary condition  and 
\cite{Sa-Va} p. 9 for Neumann-like boundary condition), gives that 
\begin{equation}\label{WM0} 
\left \{ \begin{array}{l}  
 N(\lambda , -\Delta_{A,D}^{M_0^0})= (4\pi)^{-1} |M_0^0| 
 \lambda  + \bf{O}( \sqrt \lambda)\\ 
N(\lambda , -\Delta_{A,N}^{M_0^0})= (4\pi)^{-1} |M_0^0| 
 \lambda  + \bf{O}( \sqrt \lambda)\\ 
\end{array} \right \} \; .\end{equation} 
The asymptotic formula for $ N(\lambda, -\Delta_{A,N}^{M_j})\; ,$ 
\begin{equation}\label{Neumann} N(\lambda, -\Delta_{A,N}^{M_j})\; =\; \lambda \frac{|M_j |}{4\pi} 
\; +\; {\bf O}(\sqrt{\lambda } \ln \lambda )\; , 
\end{equation} 
is  obtained as for the Dirichlet case  
(\ref{Dem}) (with $M=M_j\; ),$ by 
noticing that   
$ N(\lambda, P_{\ell ,D}) \leq  N(\lambda, P_{\ell ,N}) 
\leq N(\lambda, P_{\ell ,D}) +1\; ,$ where $  P_{\ell ,D}$ 
and $P_{\ell ,N}$ are Dirichlet and Neumann 
operators on a half-line $I=]\alpha^2 , +\infty [\; ,$ 
associated to the same differential Sch\"odinger 
operator 
$\Di P_\ell =D_t^2 +\frac{1}{4} + (e^t \frac{(\ell + \xi )}{L} \pm b)^2\; .$

We get (\ref{De}) from (\ref{Dem}) 
with $M=M_j\; ,$  (\ref{Neumann}), (for any $j=1,\ldots ,J)\; ,$ 
(\ref{Delll}) and  (\ref{WM0}).$\square $

\begin{remark}  Theorem \ref{ThD} still holds
if the metric of $\mM $ is modified in a compact set.

When $A=0\; , \ -\Delta =-\Delta_0$ has embedded eigenvalues 
in its essential spectrum, 
$(sp_{ess} (-\Delta )=[\frac{1}{4} , +\infty [ )\; .$ 
If $N_{ess}(\lambda , -\Delta )$ denotes the number 
of these eigenvalues in $[\frac{1}{4} , \lambda [\; ,$ 
then it is well known that one has an upper bound  
$\Di  N_{ess}(\lambda , -\Delta )\; \leq \; \lambda \frac{|{\bf M}|}{4\pi }\; ;$  
see \cite{Col1} and \cite{Hej}  for the history and related improvement 
of the upper bound.\\ 
Recently \cite{Mul} established a sharp asymptotic formula, 
similar to our case, 
$$ \Di  N_{ess}(\lambda , -\Delta )\; = \; \lambda \frac{|{\bf M}|}{4\pi }\; +\; {\bf O}(\sqrt{\lambda} \ln \lambda )\; ,$$
for some particular ${\bf M}\; .$ 
\end{remark}

\noindent 
{\bf Acknowledgement.} 

{\it {We are grateful to Yves Colin de Verdi\`ere 
for his useful comments and for pointing out the results 
for embedded eigenvalues of}}  $-\Delta \; .$

\end{document}